\documentclass[3p,times,twocolumn]{elsarticle}

\usepackage{ecrc}

\volume{00}

\firstpage{1}

\journalname{Nuclear and Particle Physics Proceedings}

\runauth{}

\jid{nppp}

\jnltitlelogo{Nuclear and Particle Physics Proceedings}

\usepackage{amssymb,amsthm}
\usepackage{mathtools}
\usepackage{mathrsfs}
\usepackage{bm}
\usepackage{slashed}	
\usepackage{soul}
\usepackage{hyperref}
\usepackage{lineno}
\biboptions{sort&compress}
\usepackage{verbatim}
\usepackage{fullpage}
\usepackage{graphicx}
\usepackage{tikz}
\usepackage{subfig} 	
\usepackage{relsize}	
\usepackage{float}
\usepackage{color}
\usepackage{xcolor}
\usepackage{array}
\usepackage{booktabs}
\usepackage{arydshln}
\usepackage{lscape}
\usepackage{multirow}

\def\dd{{\mathrm{d}}}

\newcommand{\dif}{\mathrm{d}}

\begin{document}

\begin{frontmatter}



\dochead{}

\title{Charmonium spectrum and diffractive production in a light-front Hamiltonian approach}


\author{Guangyao Chen\corref{cor1}}
\ead{gchen@iastate.edu}

\author{Yang~Li\corref{ca2}}
\ead{leeyoung@iastate.edu}

\author{Pieter~Maris\corref{ca3}}
\ead{pmaris@iastate.edu}

\author{Kirill~Tuchin\corref{ca4}}
\ead{tuchin@iastate.edu}

\author{James~P.~Vary\corref{ca5}}
\ead{jvary@iastate.edu}

\cortext[cor1]{Corresponding author}

\address{Department of Physics and Astronomy, Iowa State University, Ames, IA 50011, USA}

\begin{abstract}
We study exclusive charmonium production in diffractive deep inelastic scattering and ultra-peripheral heavy-ion collisions within the dipole picture. The mass spectrum and light-front wavefunctions of charmonium are obtained from the basis light-front quantization approach, using the one-gluon exchange interaction plus a confining potential inspired by light-front holography. We apply these light-front wavefunctions to exclusive charmonium production. The resulting cross sections are in reasonable agreement with electron-proton collision data at HERA and ultra-peripheral nucleus collision measurements at RHIC and LHC. The charmonium cross-section has model dependence on the dipole model. We observe that the cross-section ratio of excited states to the ground state has a weaker dependence than the cross-section itself. We suggest that measurements of excited states of heavy quarkonium production in future electron-ion collision experiments will impose rigorous constraints on heavy quarkonium light-front wavefunctions, thus improving our understanding of meson structure, which eventually will help us develop a precise description of the gluon distribution function in the small-$x$ regime.
\end{abstract}

\begin{keyword}

charmonium \sep light front \sep meson production \sep dipole model

\end{keyword}

\end{frontmatter}


\section{Introduction}
\label{sec:intro}

Exclusive vector meson production processes are valuable probes of hadron structures \cite{Gribov:1984tu} and provide insights to QuantumChromodynamics (QCD) in the high energy limit, where saturation dominates the gluon dynamics \cite{Gribov:1984tu,JalilianMarian:1996xn,Gelis:2010nm}. Models incorporating the saturation physics have been very successful in describing high precision electron-proton collision data collected at the Hadron-Electron Ring Accelerator (HERA) \citep{Golec-Biernat:1998js,Levin:2000mv,Gotsman:2001ne,Kowalski:2003hm,Iancu:2003ge}. 

The diffractive DIS process can be effectively approximated by the scattering of a color dipole, a quark-antiquark pair, from the proton \cite{Mueller:1989st,Nikolaev:1990ja}. The so called dipole picture has been very successful in explaining both exclusive and diffractive HERA measurements in the high-energy limit \citep{Kowalski:2006hc, Marquet:2007nf}, by employing some phenomenological vector meson light-front wavefunction (LFWF) \citep{Kowalski:2003hm,Kowalski:2006hc}. Such phenomenological models contain free parameters that weaken the predictive power of the diffractive heavy quarkonium production process. 

Recently, a new description of heavy quarkonium system has emerged \cite{Li:2015zda,Li:2016wwu} within the basis light-front quantization (BLFQ) approach \cite{Vary:2009gt, Honkanen:2010rc, Vary:2016emi}. The mass spectra for charmonium and bottomonium are obtained by diagonalizing a Hamiltonian within the BLFQ framework, with the one-gluon exchange interaction and a confining potential inspired by light-front holography \cite{Li:2015zda}. The successful applications of the BLFQ formalism to the electron anomalous magnetic moment \cite{Honkanen:2010rc,Zhao:2014xaa}, and to the positronium system \cite{Wiecki:2014ola,Adhikari:2016idg} have paved the way for the study of the heavy quarkonium system. The LFWFs from the BLFQ approach, which arise from successful fits to the heavy quarkonia mass spectroscopy, show success in applications to decay constants and to additional observables such as charge form factors. Here we report predictions of the LFWFs obtained from the BLFQ approach and compare with selected experiment data on diffractive charmonium production, which were discussed in detail in Ref.~\citep{Chen:2016dlk}. 

\begin{figure}
 \centering 
\includegraphics[width=.45\textwidth]{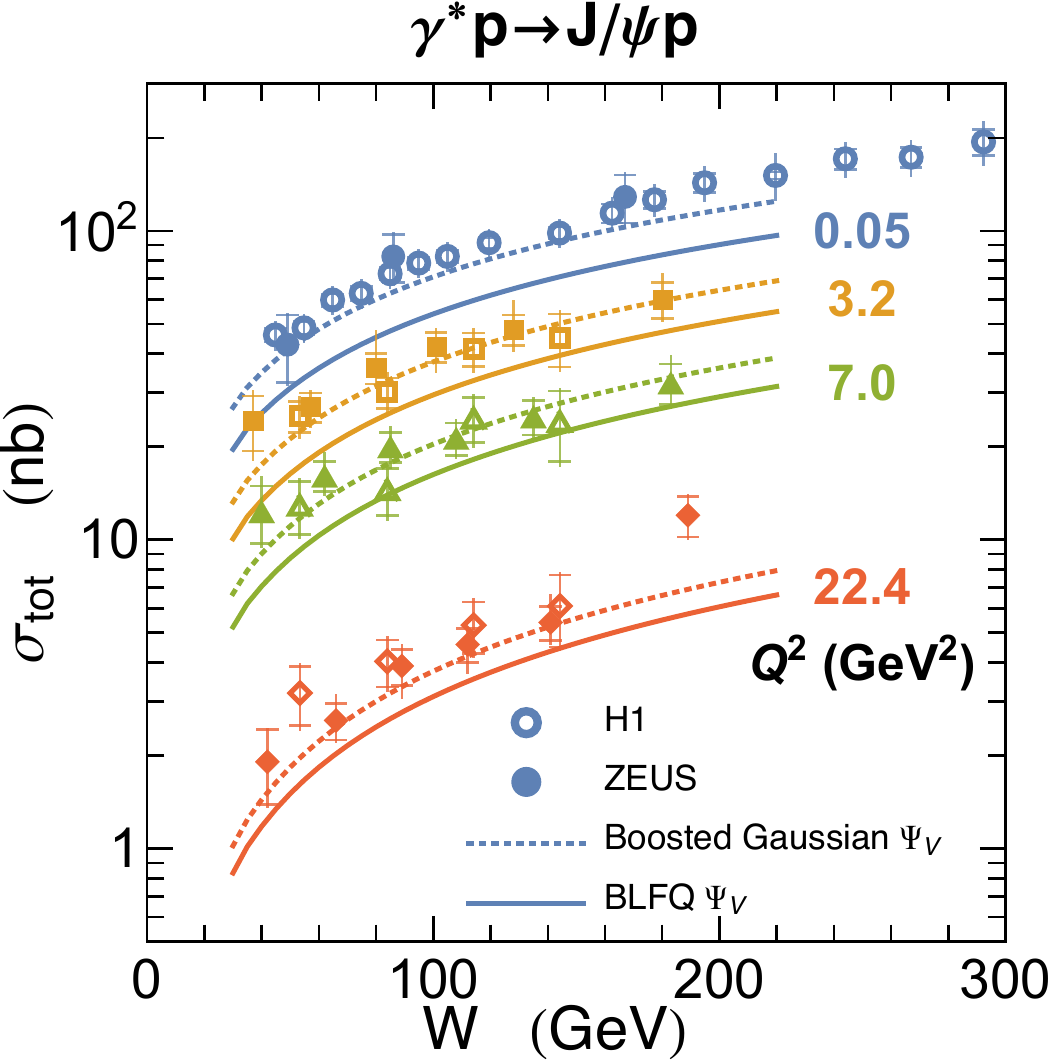}
\caption{Predictions of the BLFQ LFWF (solid curves) and the boosted Gaussian LFWF (dashed curves) compared with the HERA experimental data of total $J/\Psi$ cross section for different values of $Q^2$ and $W$ \cite{Chekanov:2004mw,Aktas:2005xu}. The inner bars indicate the statistical uncertainties; the outer bars are the statistical and systematic uncertainties added in quadrature.
}
\label{fig:hera1}
\end{figure}

\section{Theoretical framework}  
The amplitude for producing an exclusive vector meson in diffractive DIS is calculated as follows in the dipole picture \cite{Kowalski:2006hc},
\begin{align}
  \mathcal{A}^{\gamma^* p\rightarrow Ep}_{T,L} = &\mathrm{i}\,\int\!\dif^2\bm{r}\int_0^1\!\frac{\dif{z}}{4\pi}\int\!\dif^2\bm{b}\;(\Psi_{E}^{*}\Psi)_{T,L} (r,z,Q) \; \nonumber \\
  & \mathrm{e}^{-\mathrm{i}[\bm{b}-(1-z)\bm{r}]\cdot\bm{\Delta}}
  \;\frac {\dif\sigma_{q\bar q}}{\dif^2 \bm b} (x,r) \; ,
  \label{eq:newampvecm} 
\end{align}
where $Q^2$ is the virtuality of photon, $T$ and $L$ denote the transverse and longitudinal polarization of the produced vector meson, and the momentum transfer being $t=-\vec{\Delta}^2$. $\vec{r}$ is the transverse separation between the quark and antiquark and $z$ is the LF longitudinal momentum fraction carried by the quark respectively. $\vec{b}$ is the impact parameter of the dipole relative to the proton and $x$ is the Bjorken variable. $\Psi$ and $\Psi_{E}^{*}$ are LFWFs of the virtual photon and the exclusively produced vector meson respectively. The cross section is related to the amplitude as
\begin{eqnarray}
\frac{\dif \sigma^{\gamma^* p\rightarrow Ep}_{T,L}}{\dif t} = \frac{1}{16 \pi} \vert \mathcal{A}^{\gamma^* p\rightarrow Ep}_{T,L}(x,Q,\Delta) \vert^2 \; .
\end{eqnarray}
Moreover, contributions from the real part of the scattering amplitude and skewedness correction should be taken into account, see Ref.~\citep{Chen:2016dlk} for details.

We employ the impact parameter dependent saturation (bSat) model \cite{Kowalski:2003hm} and the impact parameter dependent Color Glass Condensate (bCGC) model \cite{Iancu:2003ge} for this study. We use five sets of parameters (bSat I-V) in the bSat model\cite{Kowalski:2006hc,Rezaeian:2012ji} and three sets of parameters (bCGC I-III) in the bCGC model \cite{Soyez:2007kg,Rezaeian:2013tka}. The parameters for these dipole cross section parametrizations are summarized in Tables 1 and 2 in Ref.~\citep{Chen:2016dlk}.

The heavy quarkonium mass spectrum and LFWFs are obtained by solving the eigenvalue equation of an effective light-front Hamiltonian, which combines the holographic QCD Hamiltonian \cite{Brodsky:2014yha} and the one-gluon exchange dynamics \cite{Li:2015zda},
\begin{equation}
 H_\text{eff}|\psi_h\rangle = M^2_h|\psi_h\rangle, \quad (H_\text{eff} \equiv P^+\hat P^-_\text{eff} - \vec P^2) \; .
\end{equation}
with
\begin{align}
\label{eq:Ham}
 H_\text{eff} = & \frac{\bm k_\perp^2 + m_q^2}{z(1-z)}  + \kappa_\text{con}^4 \bm \zeta_\perp^2 
 - \frac{\kappa_\text{con}^4}{4 m_q^2}\partial_z \big(z(1-z) \partial_z \big) \nonumber \\
 &-\frac{4\pi C_F \alpha_s}{Q^2} \bar u_{s}(k)\gamma_\mu u_{s'}(k') \bar v_{\bar s'}(\bar k') \gamma^\mu v_{\bar s}(\bar k)\;,
 \end{align} 
where $C_F=\frac{4}{3}$, $Q^2=-\frac{1}{2}(k'-k)^2-\frac{1}{2}(\bar k-\bar k')^2$. The strong coupling constant $\alpha_s$ is fixed, $\alpha_s(M_{c\bar c}) \simeq 0.36$ and $\alpha_s(M_{b\bar b}) \simeq 0.25$. The effective quark mass $m_q$ and the confining strength $\kappa_\text{con}$ are determined by fitting the heavy quarkonium mass spectrum to the experimental measurements. The calculated spectra agree with the experimental values within a root-mean-square deviation of around $50$~MeV for the states below the open flavor thresholds. 

\begin{figure}
 \centering 
\includegraphics[width=.45\textwidth]{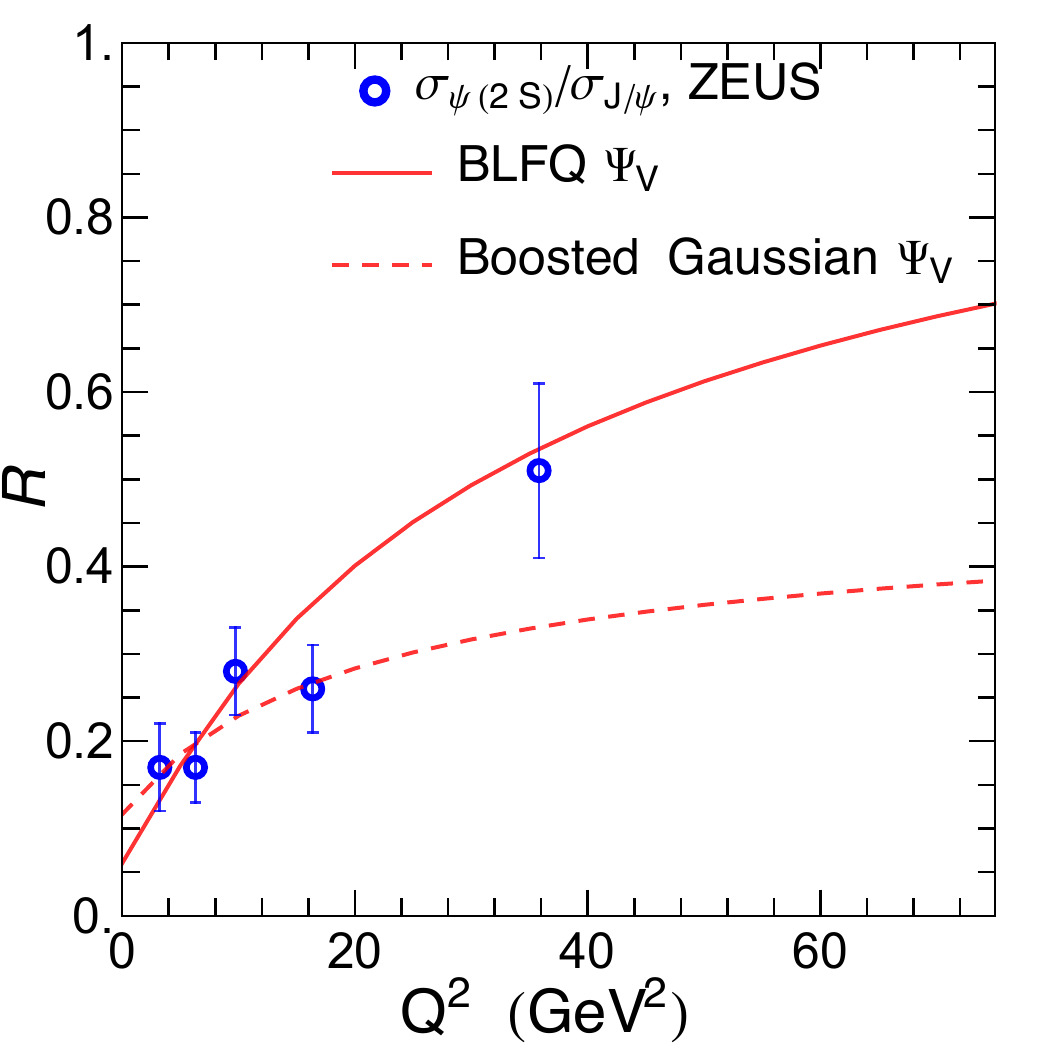}
\caption{Predictions of the BLFQ LFWF (solid curves) and the boosted Gaussian LFWF (dashed curves) compared with the HERA experimental data for the $\sigma_{\Psi(2s)}/\sigma_{J/\Psi}$ cross-section ratio as a function of $Q^2$ in electron-proton scattering \cite{Abramowicz:2016ext}. Error bars indicate the statistical uncertainties.
}
\label{fig:hera2}
\end{figure}

\begin{figure}
 \centering 
\includegraphics[width=.45\textwidth]{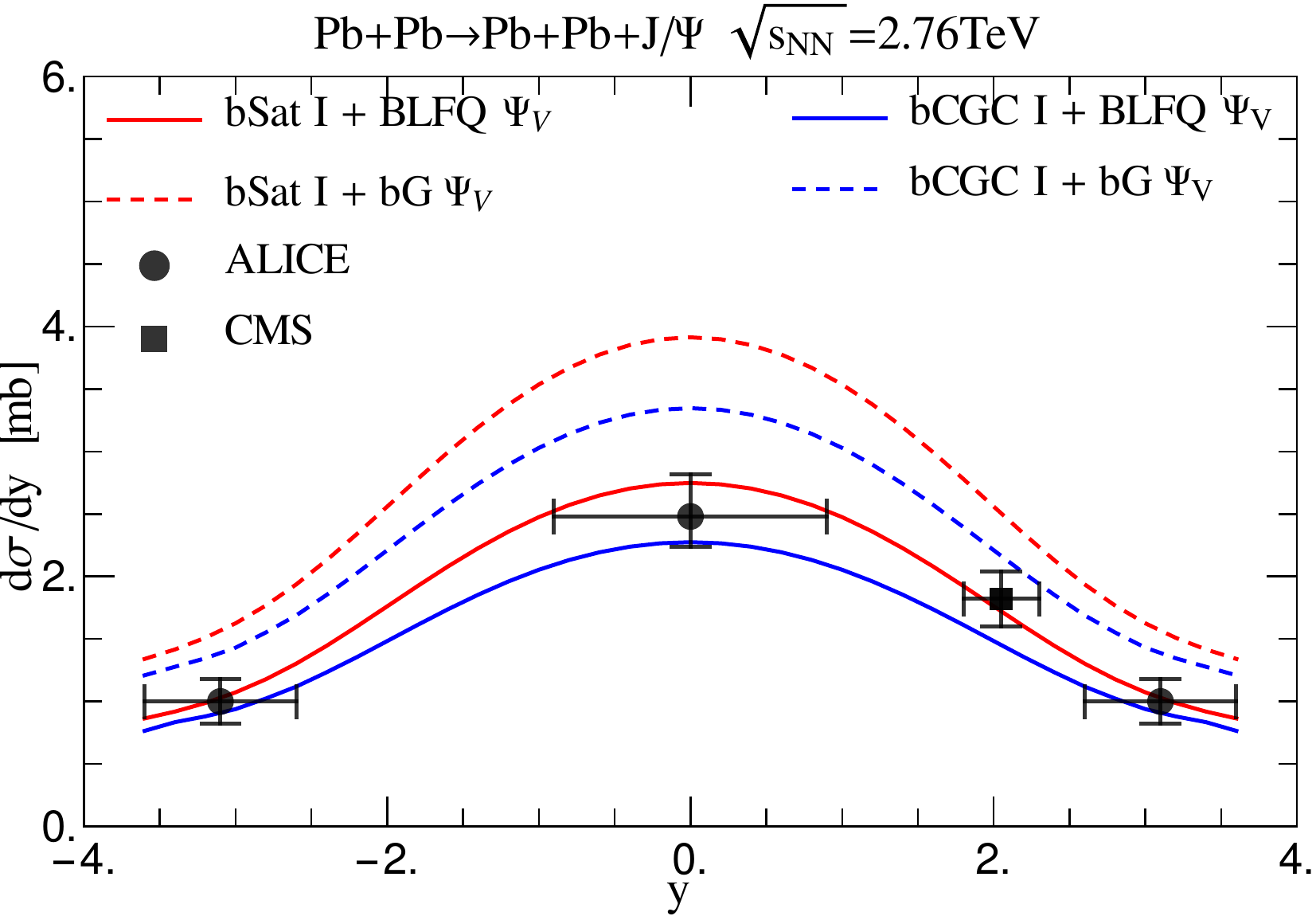}
\caption{The predictions of the BLFQ LFWF (solid curves) and the boosted Gaussian LFWF (dashed curves) for the coherent production of $J/\Psi$ production in ultra-peripheral collision at $\sqrt{s_{NN}}=2.76$~TeV, compared to the measurements by the ALICE collaboration \cite{Abbas:2013oua} and CMS collaboration \cite{Khachatryan:2016qhq} at LHC. Error bars show statistical uncertainties only.  
}
\label{fig:lhc}
\end{figure}

The heavy quarkonium LFWF from the BLFQ approach has several advantages over LFWFs from phenomenological models. First, it is constrained by a variety of observables. Second, it provides access to higher excited states without introducing additional assumptions. Moreover, it can be improved by including higher Fock sectors, e.g., the quark-antiquark-gluon sector.

\section{Charmonium production at HERA, RHIC and LHC} 
\label{sec:hera}

We study the diffractive charmonium production using the heavy quarkonium LFWF from the BLFQ approach. The resulting cross sections as a function of various kinematic variables are in reasonable agreement with electron-proton collision data at HERA and ultra-peripheral nucleus collision measurements at RHIC and LHC. We present some representative results using the BLFQ LFWFs of heavy quarkonium and compare our predictions with the predictions of boosted Gaussian LFWFs.

To make comparison with HERA measurements \cite{Chekanov:2004mw,Aktas:2005xu}, we employ the bSat II dipole model in Table 1 in Ref.~\citep{Chen:2016dlk}. In Fig.~\ref{fig:hera1}, solid and dashed curves are predictions of the total $J/\Psi$ cross section as a function of various values of $Q^2$ and $W$ by the BLFQ and the boosted Gaussian LFWFs respectively. Except when $Q^2$ is very small, the predictions of the BLFQ LFWF are slightly lower than the experimental measurements. Note that the theoretical uncertainty in the dipole model is large at small $Q^2$. Fig.~\ref{fig:hera2} shows the predictions of the BLFQ LFWFs (solid curve) and the boosted Gaussian LFWFs (dashed curve) for the $\sigma_{\Psi(2s)}/\sigma_{J/\Psi}$ cross-section ratio as a function of $Q^2$ in electron-proton scattering measured at HERA \cite{Abramowicz:2016ext}. The BLFQ LFWFs provide a better fit to the data at larger $Q^2$ region.  

Using the BLFQ LFWF, the bSat I dipole parametrization and the bCGC I dipole parametrization yield a prediction of $\dd \sigma/\dd y =59.9$~$\mu$b and $\dd \sigma/\dd y = 52.6$~$\mu$b respectively for the coherent $J/\Psi$ production at mid-rapidity with two gold nuclei colliding at $\sqrt{s_{\text{NN}}}=200$~GeV at RHIC. Both results are consistent with latest data  $\dd \sigma/\dd y = 45.6 \pm 13.3$ (stat) $\pm 5.9$ (sys) $\mu$b \cite{Afanasiev:2009hy}.

In Fig.~\ref{fig:lhc}, solid curves show the prediction of the BLFQ $J/\Psi$ LFWF using bSat I (red) and bCGC I (blue) dipole model parametrizations for coherent production of $J/\Psi$ at mid-rapidity in ultra-peripheral Pb-Pb collisions at $\sqrt{s_{NN}}=2.76$~TeV \cite{Abbas:2013oua,Khachatryan:2016qhq}. We find both of these results are within the statistical uncertainty of the experimental data. In contrast, dashed curves are the predictions of the boosted Gaussian LFWF of $J/\Psi$ using bSat I (red) and bCGC I (blue) dipole model parametrizations. We find both of these overshoot the data. Both the BLFQ LFWF and boosted Gaussian LFWF underestimate the production of $\Psi(2s)$ in ultra-peripheral Pb-Pb collisions at $\sqrt{s_{NN}}=2.76$~TeV  \cite{Adam:2015sia}. 

\section{Dipole model dependence}

\begin{figure}
 \centering 
\includegraphics[width=.45\textwidth]{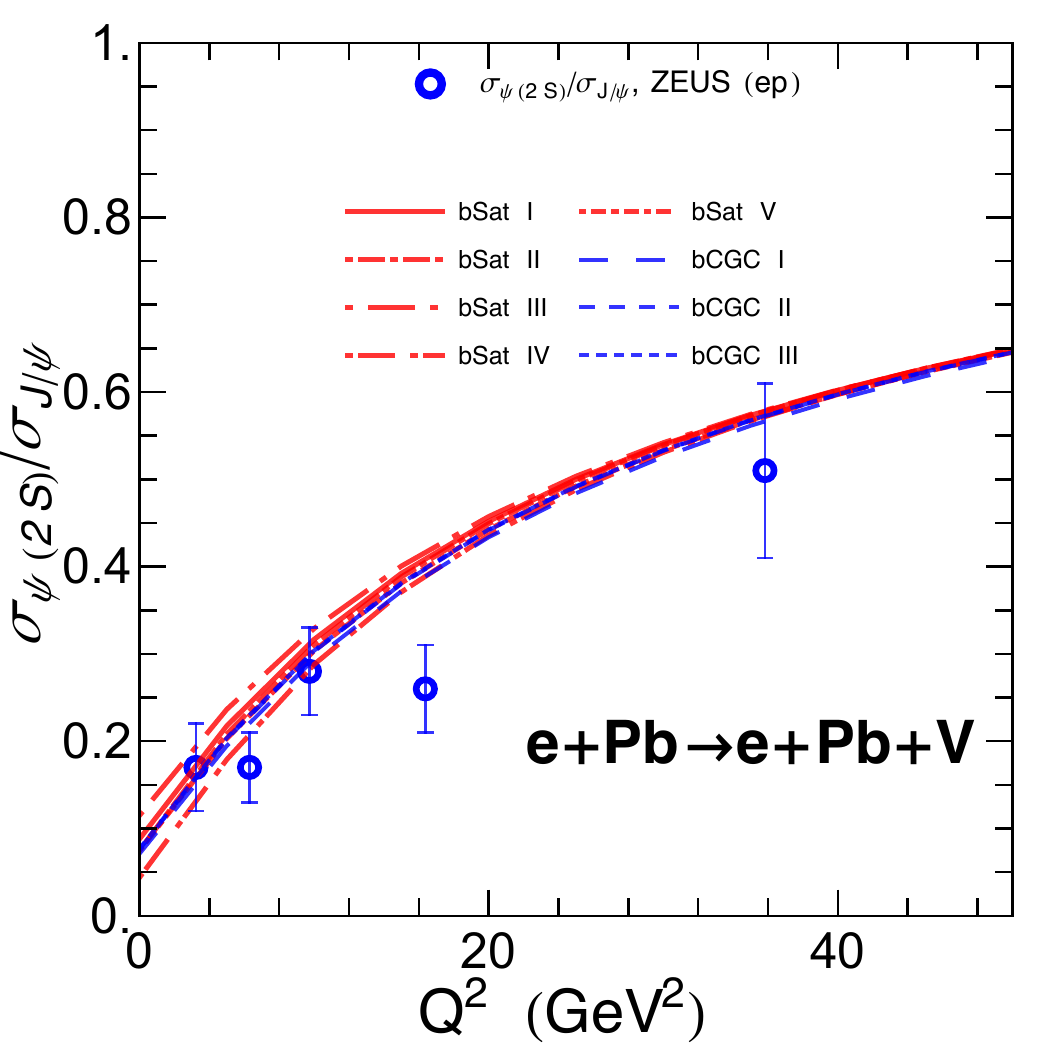}
\caption{The cross-section ratio $\sigma_{\Psi(2s)}/\sigma_{J/\Psi}$ as a function of $Q^2$ predicted by BLFQ LFWF and various dipole cross section parametrizations in Tables 1 and 2 in Ref.~\cite{Chen:2016dlk} for coherent charmonium production in electron-lead collisions. The experiment data points are measurements by ZEUS collaboration \cite{Abramowicz:2016ext} in electron-proton scattering at HERA.}
\label{fig:model}
\end{figure}

The uncertainties associated with the heavy quarkonium LFWFs and the dipole cross section both contribute to uncertainties in the results for heavy quarkonium production in the dipole picture. On the other hand, the uncertainties from the dipole cross section parametrization may be correlated for the calculation of different states of the same quarkonium system, for example, $J/\Psi$ and $\Psi(2s)$ states of charmonium. It could lead to weaker dependence on the dipole model for the calculation of the cross-section ratio of higher excited states over the ground state than the calculation of the cross section itself. We calculate the ratio of the $\Psi(2s)$ cross section to the $J/\Psi$ cross section as a function of $Q^2$ predicted by the BLFQ LFWF and various dipole cross section parametrizations in Tables 1 and 2 in Ref.~\cite{Chen:2016dlk} for electron-proton collisions and electron-ion collision as well. We observe that the cross-section ratio exhibits insignificant dependence on dipole models, especially in the large $Q^2$ regime. The case for electron-lead collisions is shown in Fig.~\ref{fig:model}.

\section{Conclusions}
We study diffractive charmonium production with BLFQ light-front wavefunctions within the dipole model. It has been shown that the effective Hamiltonian incorporating one-gluon exchange dynamics from QCD Lagrangian and effective confining potential inspired by soft-wall light-front holographic QCD reproduces the heavy quarkonium spectrum \citep{Li:2015zda}. We found the resulting charmonium LFWF gives compatible descriptions of diffractive $J/\Psi$ and $\Psi(2s)$ production data at HERA, RHIC and LHC \citep{Chen:2016dlk}. The cross-section ratio of $\sigma_{\Psi(2s)}/\sigma_{J/\Psi}$ as a function of $Q^2$ shows a weak dependence on the dipole model, which could lead to a reduction of theoretical uncertainties associated with the structure of heavy quarkonium at future electron-ion collision experiments \cite{AbelleiraFernandez:2012cc,Accardi:2012qut}, by measuring the cross-section ratios of the higher excited states to the ground state. The gluon distribution in the saturation regime could be extracted efficiently through diffractive production processes with accurate heavy quarkonium LFWFs. 

Our preliminary results show that, the predictions using BLFQ LFWFs lead to reasonable agreement with diffractive charmonium and bottomonium production in UPC at LHC through proton-proton, proton-lead and lead-lead collisions at various energy scales \citep{Chen:2016upc}. The incoherent production of charmonium states and bottomonium states at the LHC and the EIC are under investigation. Future improvement includes increasing the Fock sectors for the heavy quarkonium system and to reduce the theoretical uncertainties associated with non-perturbative effects of the dipole model approximation \cite{Vary:2016wrd}.

\section*{Acknowledgments} 
We thank X. Zhao, P. Wiecki and Y. Xie for valuable discussions. 
We thank N. Kovalchuk for providing us the experimental data for $\Psi(2s)$ measurement.
This work was supported in part by the Department of Energy under
Grant Nos. DE-FG02-87ER40371 and DESC0008485 (SciDAC-3/NUCLEI). We acknowledge computational resources provided by the National Energy Research Scientific Computing Center (NERSC) under Contract No. DE-AC02-05CH11231.




\nocite{*}
\bibliographystyle{elsarticle-num}
\bibliography{jos}



\end{document}